# Sampling the proteome by emerging single-molecule and mass spectrometry methods

Michael J. MacCoss[1,#], Javier Alfaro[2,3,#], Meni Wanunu[4], Danielle A. Faivre[5], Christine C. Wu[6], & Nikolai Slavov[7,8,#]

1. Department of Genome Sciences, University of Washington, Seattle, WA 98117, USA 0000-0003-1853-0256   maccoss@uw.edu
2. International Centre for Cancer Vaccine Science, University of Gdańsk, Gdańsk, Poland. javier.alfaro@proteogenomics.ca
3. Department of Biochemistry and Microbiology, University of Victoria, Victoria, BC, Canada
4. Department of Physics, Northeastern University, Boston, MA 02115, USA 0000-0002-9837-0004 wanunu@neu.edu
5. Department of Genome Sciences, University of Washington, Seattle, WA 98117, USA   dfaivre@uw.edu
6. Department of Genome Sciences, University of Washington, Seattle, WA 98117, USA chriscwu@uw.edu
7. Departments of Bioengineering, Biology, Chemistry and Chemical Biology, Single Cell Proteomics Center, and Barnett Institute, Northeastern University, Boston, MA 02115, USA;  0000-0003-2035-1820 nslavov@northeastern.edu
8. Parallel Squared Technology Institute, Watertown, MA 02472, USA
   # Contributed Equally

## Abstract

Mammalian cells have about 30,000-fold more protein molecules than mRNA molecules. This larger number of molecules and the associated larger dynamic range have major implications in the development of proteomics technologies. We examine these implications for both liquid chromatography-tandem mass spectrometry (LC-MS/MS) and single-molecule counting and provide estimates on how many molecules are routinely measured in proteomics experiments by LC-MS/MS. We review strategies that have been helpful for counting billions of protein molecules by LC-MS/MS and suggest that these strategies can benefit single-molecule methods, especially in mitigating the challenges of the wide dynamic range of the proteome. We also examine the theoretical possibilities for scaling up single-molecule and mass spectrometry proteomics approaches to quantifying the billions of protein molecules that make up the proteomes of our cells.

## Introduction

The ubiquitous roles of proteins in biomedicine are well appreciated and have motivated technologies seeking to advance the *sensitivity and throughput* of quantitative protein analysis. While proteomic technologies may use different approaches, they face similar challenges, such as quantifying proteins of vastly different abundances, some present in only a few copies and some present in tens of millions of copies per typical mammalian cell. This wide dynamic range poses a substantial challenge for investigating proteome biology.

Mass spectrometry (MS) has powered proteomics from the first demonstration of peptide sequencing using MS in the 1970s[1–3]. Since then, milestones in MS-based proteomics have included *de novo* sequencing entire proteins in the late 1980s[4–7], soft ionization by electrospray[8], automated spectral interpretation[9], multiplexing the acquisition of spectra on different peptides using data independent acquisition[10], multiplexing the acquisition of different samples using tandem mass tags[11], and quantifying thousands of proteins in single human cells[12,13]. Together, the steady growth in the rate of protein identification using MS has been reminiscent of Moore's law, resulting in about 1,250-fold higher throughput: from about 20 protein data points per hour in 2001[14] to about 25,000 protein data points per hour achieved by plexDIA[15]. This increased throughput has been critical for addressing challenges in biomedical research[16]. It also highlights the power of experimental strategies and technological progress to tackle the immense demands of proteomics in terms of quantity and dynamic range that is required for thorough analysis, given the large number of proteins of widely varying concentrations in a cell.

More recently, non-MS methods have made exciting steps towards identifying and potentially sequencing single polypeptide molecules[17–19]. Conceptually, these methods aim to adapt flow-cell and nanopore methods developed for nucleic acid analysis for protein analysis. Flow-cell based methods include highly parallel single-molecule N-terminal peptide sequencing methods based on either Edman degradation[20] or amino peptidases[21]. Another approach aims to use degenerate affinity reagents to recognize individual protein molecules separated spatially in a flow cell[22,23]. Other groups are working to adapt nanopore sequencing to peptides and proteins[24,25]. Most of these methods aim to detect a subset of the amino acids within a polypeptide sequence, which provides a fingerprint, or a constraint, on choosing a sequence among the known protein coding gene products from the genome. While these methods have yet to be applied to biologically derived protein mixtures, they have generated significant enthusiasm within the scientific community as a complement to MS analysis[17].

These developments have motivated renewed interest and investment in advancing proteomics technologies, as reflected in private funding[17] and in recent National Human Genome Research Institute (NHGRI) funding opportunities aimed at accelerating the development of technologies for single-molecule sequencing and single-cell proteome analysis. Because there is excitement for new emerging single-molecule counting methods for proteomics, we felt it was timely to provide a perspective that compares strategies used by the current state-of-the-art proteomics methods based on liquid chromatography-tandem mass spectrometry (LC-MS/MS) to the new and emerging counting-based-methods that enabled complete and accurate transcriptome sequencing. We hope our opinion will provide benchmarks and directions for the technological breakthroughs that need to be achieved for single molecule protein/peptide counting to achieve parity and complement the capabilities of LC-MS/MS based proteomics methods. How many molecules need to be counted? How extreme is the dynamic range problem? Do current solutions for handling the dynamic range problem limit the sequence coverage of the proteome? What will new technologies need to accomplish to reach parity with LC-MS/MS, and how will these technologies complement one another? What can these emerging technologies learn from LC-MS/MS based proteomics? These are the questions we aim to address.

## How many molecules need to be counted?

Many of the challenges for accurate and sensitive protein quantification are shared by all proteomics methods, such as the quantification over a wide dynamic range. Indeed, a typical mammalian cell contains billions of protein molecules but less than half a million RNA molecules[26], Figure 1a. Some proteins are present at hundreds of copies per cell while others (e.g., histones) at tens of millions of copies per cell, resulting in about $10^6$ dynamic range[27]. The range of protein abundances is even larger for body fluids, such as plasma where protein abundances may differ by $10^{10}$, e.g., between albumin and IL-6[28]. This wide range of abundance is a fundamental challenge for proteomics because the presence of abundant proteins make it rare to count molecules from low abundant proteins, such as having to count billions of albumin molecules before having a chance to detect a single IL-6 molecule. This means that an emphasis on highly sensitive single-molecule approaches that have been used successfully to quantify the transcriptome, which spans about $10^3$ dynamic range, face major challenges in scaling to quantify the proteome[25].

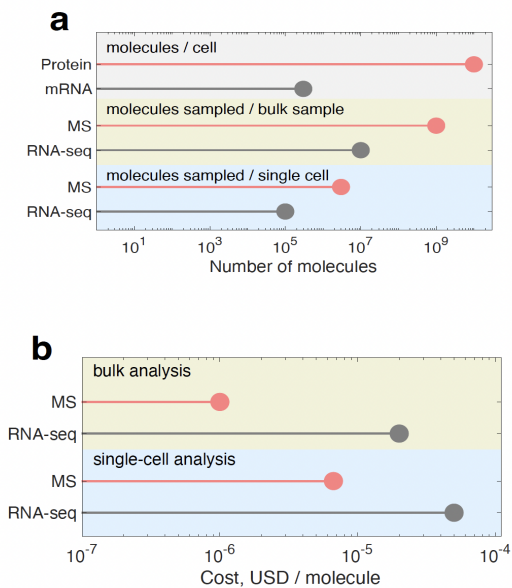

**Figure 1 | Overview of RNA and protein statistics.** **a**, A representative human cell, such a fibroblast, has billions of protein molecules compared to merely hundreds of thousands of RNA molecules[29]. Accordingly, MS analysis samples more protein molecules per sample than the RNA molecules sampled by RNA-seq. **b**, Estimated cost per molecule for MS and RNA-seq. The single-cell estimates are based on published numbers for unique molecular identifiers per single cell analyzed by Smart-seq3[30] and number of protein molecules counted by plexDIA[15].

A typical mammalian cell, i.e., a HeLa cell with a volume of ~3,000 µm³, contains about 300,000 mRNA molecules[29] and about 10,000,000,000 protein molecules[26], Figure 1a. The cell is a crowded mesh of proteins, with a typical density of 3 million protein molecules per cubic micrometer. Even a yeast cell with a volume of ~30 µm³ contains ~100 million molecules. This protein density estimate has been supported independently using molecular measurement based on MS, as well as fluorescence microscopy using green fluorescent protein. Given these different independent measurements, it is estimated that the typical HeLa cell contains at least ~3-5 billion proteins per cell and others like macrophages (5,000 µm³) and cardiomyocytes (15,000 µm³) will contain substantially more. Because of this range in volume, we used ~10 billion proteins per cell in our calculations. These estimates of the relative abundance ratio of mRNA to protein molecules has the direct consequence of requiring about 30,000-fold more counts to characterize the protein molecules at an analogous coverage of what has been achieved with the transcriptome, Figure 1. Given the potential need to count a large number of protein molecules, we next explore the feasibility of achieving the required scale at affordable cost using estimates for cost per molecule. This factor is important, but it must be considered in the context of many other factors, such as the ability to sample large numbers of diverse sequences and to multiplex efficiently.

## How much do single molecule counting methods cost?

While single-molecule protein counting approaches are yet to report the analysis of a complex protein mixtures, we believe that with time and resources the efforts of reading peptide sequences in a spatially parallelized format will be successful[17]. Without knowing what the capabilities and limitations are for these emerging protein and peptide sequencing methods – we make the optimistic assumption that these methods will be able to achieve sequencing counts of polypeptides on par with what state-of-the-art Illumina sequencing can achieve currently with oligonucleotides. Thus, we use single-molecule RNA sequencing by Illumina as a proxy to represent single-molecule protein counting approaches, Fig. 1b. To estimate the cost for current advanced technologies, we use the estimate of $10,000 for sequencing 4 billion reads by Illumina NovaSeq over ~2-days and $500 for performing a 2-hour quantitative LC-MS/MS analysis. These costs were chosen as conservative estimates based on inquiries from several academic core facilities. While academic research laboratories may achieve lower costs, these prices represent objective estimates for widely accessible services. The results in Fig. 1b indicate that the cost per protein molecule analyzed by LC-MS/MS is lower than the cost of DNA molecule sequenced by Illumina. This indicates that single-molecule DNA sequencing has not yet achieved a cost that would enable counting of sufficient numbers of molecules to achieve affordable and comprehensive quantification of mammalian proteomes.

## Counting ions by LC-MS/MS

Traditionally, the MS proteomics field reports lists of peptides detected and the proteins they are derived from. As peptides elute off the column, the instrument counts large numbers of peptide ions based on their mass-to-charge (m/z) ratio, independently of their sequence identification (**Figure 2a**). The abundance of each analyte is often determined from a background subtracted peak area of the extracted ion chromatogram(s). Depending on the method used, the peak area can be obtained from the unfragmented MS1 spectra or from tandem mass spectra (MS/MS or MS2) collected using methods like data independent acquisition. The peak area is derived from the detector ion current, either from the flow of ions to an electron multiplier[31] or the generation of an image current in a Fourier transform mass analyzer[32]. The current is a measure of the number of ions (charged molecules) counted, normalized by the amount of time spent sampling the signal. The measured signal is proportional to ions/second, and thus, it can be converted into a number of counted ions and compared directly with single molecule counting methods[12,33,34].

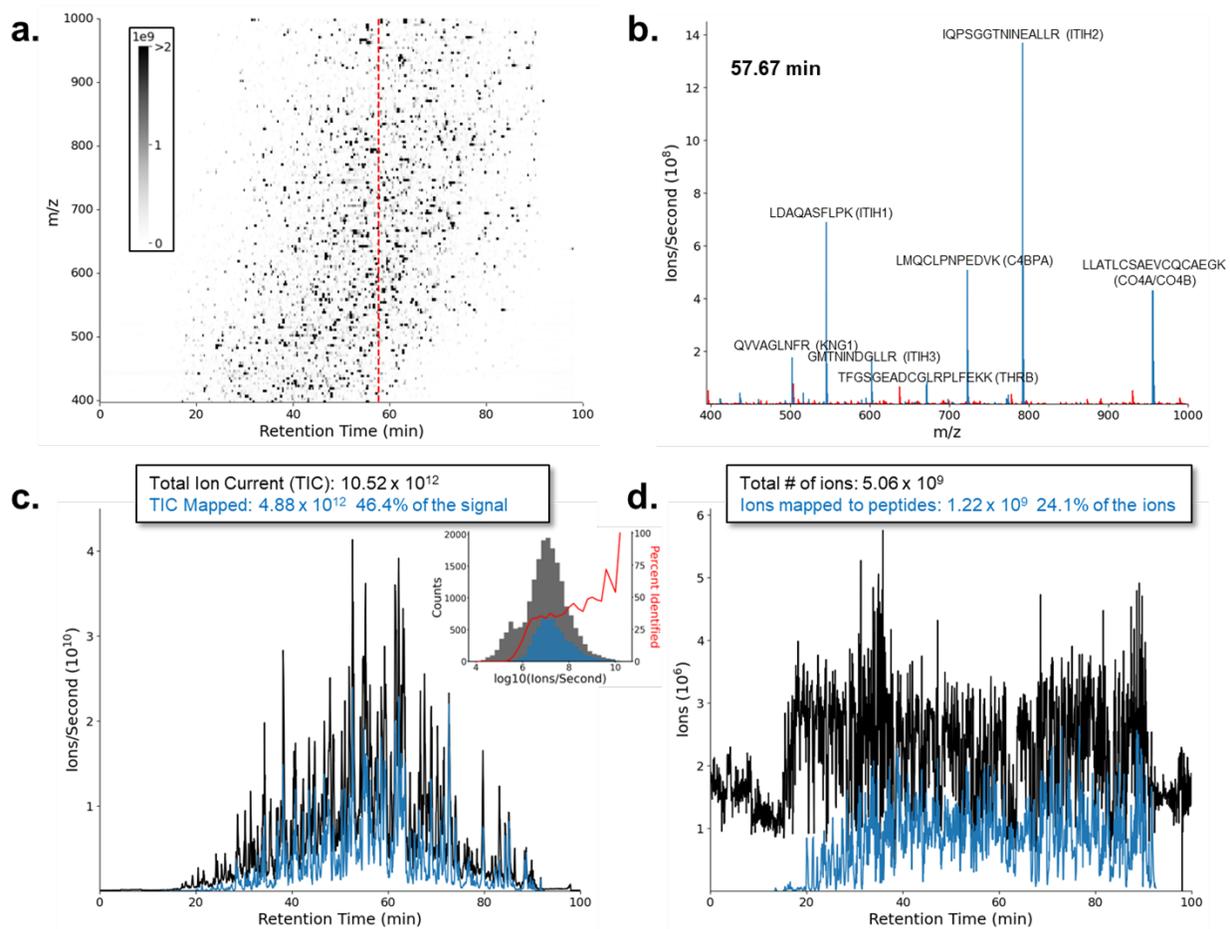

**Figure 2 | A liquid chromatography-mass spectrometry experiment can count billions of peptide ions within 90 min.** Signal from the MS1 spectra of an LC-MS run of enriched extracellular vesicles from human plasma using data independent acquisition of an ThermoFisher Eclipse Tribrid. **a**, A 2-dimensional ion map of the MS1 peptide signal separated in both retention time (RT) and *m/z* dimensions. The red dashed line indicates the location of the spectrum in (b). **b**, Selection of a single MS1 spectrum collected at 57.67 min where blue m/z values have been assigned a peptide sequence and red m/z values are unassigned in the analysis. **c**, A total ion current (TIC) plot of the signal intensity from (a) at all time points. The TIC signal is plotted in black, and the blue represents the fraction of the MS1 signal (e.g., in b) that has been confidently assigned to peptide sequences. The y-axis represents an approximation of counts (ions per second). The insert is a histogram counting distinct molecular entities (features) for different measured intensities. The gray bars of the insert represent all molecular features and the blue represents those assigned a peptide label. The data were only analyzed for unmodified and fully tryptic peptides in the canonical human fasta. **d**, Represents the same data plotted in (c) but with the y-axis of each spectrum adjusted to an estimate of ions by multiplying the counts by the Orbitrap fill time. The variable fill times allow peptides with relatively low abundance near 20-30 min to be measured with a similar number of ions as the most abundant peptides in the analysis. The result is billions of peptide ions counted within just 90 minutes. Data available at: https://panoramaweb.org/Single_Molecule_Counting.url under PXD035637. Code available at: https://github.com/uw-maccosslab/single_molecule_counting.

LC-MS/MS methods can improve the sensitivity to low-abundance analytes by changing the time spent sampling the signal (aka dwell time, integration time, or injection time). In some MS instruments, such as ion traps, the time spent sampling ions changes dynamically depending on the signal at that time[35]. This dynamic adjustment of the injection time, known as automatic gain control (AGC), provides an ideal ion population for the MS measurement (**Figure 2b**). However, an added benefit of AGC is that it enables the instrument to spend less time on abundant molecular species but scale the current into a larger quantity while maintaining quantitative linearity. Likewise, it enables spending more time on less abundant peptides to enable the measurement of the weaker signal; This increases the dynamic range and the total number of ions identified (**Figure 2c**). Dividing each spectrum intensity by the time taken to acquire the spectrum gives a normalized signal for each spectrum that is analogous to normalizing the counts obtained between flow cells in a single-molecule counting experiment[36,37].

LC-MS has a much greater dynamic range than would be expected from simply counting the billions of ions and assigning the counts to peptides. This increase in dynamic range arises because LC-MS first chromatographically separates peptides based on their physical properties so that peptides of the same sequence are measured together (Figure 3). This strategy of counting the same peptide sequences together to provide a quantity is effectively a compression scheme for counting molecules. Additionally, using gas-phase methods, MS can further improve the dynamic range by measuring the m/z of all peptides and fragments with the same values together. Thus, the effect of highly abundant peptides on the measurement of lowly abundant peptides is minimized because they are measured separately and in some experiment types, separate trap fills (i.e. analogous to measuring abundant transcripts in different flow cells from low abundance transcripts). In fact, over the years the mass spectrometry community has capitalized on this strategy to improve the detection and precision of low abundant molecules[10,38–40] in the presence of analytes with much greater abundance. Because the timescale of this measurement is fast (sub-second) MS can analyze such compressed groups of ions (~10 to $1\times10^6$ ion copies at a time) tens of thousands of times per hour.

Summing up with an example, a 90 min LC-MS/MS analysis of peptides in plasma frequently measures $3\times10^9$ ions from just the unfragmented MS1 signal. Yet, this frequently represents only peptides from ~350-450 proteins because the dynamic range of the plasma proteome is notoriously large[41]. Thus, if plasma is analyzed using a single flow cell with 1 million single molecule "reads", ~950,000 of those reads would be of the 12 most abundant proteins,[28] leaving only 50,000 (or 5%) of the remaining reads to quantify the rest of the proteins in the sample.

The dynamic range of plasma can be mitigated by depleting the most abundant proteins by immunoaffinity subtraction chromatography[42]. Such chromatography frequently removes 14 of the most abundant proteins in human plasma (e.g., albumin, IgG, antitrypsin, IgA, transferrin, haptoglobin, fibrinogen, alpha2-macroglobulin, alpha1-acid glycoprotein, IgM, apolipoprotein A1, apolipoprotein A2, complement C3 and transthyretin). Depletion increases the number of detected proteins, but unfortunately these affinity columns are species specific and thus are largely limited for use with human samples. These columns also capture the entire complex and binding proteins of the target antigens – removing unintended proteins. For example, albumin binds as many as 35 proteins and the albuminome itself has been proposed as a plausible plasma subfraction of biomedical interest[43]. It is well known that patients with cancer make autoantibodies to known cancer biomarkers[44] (e.g. thyroglobulin, MUC16 (CA125), and PSA) which complicate their analysis using immunoaffinity methods[45] and further, depletion of IgGs can remove these biomarkers. Depletion of apolipoprotein A1, will also deplete HDL particles[46], a promising plasma sub proteome for the diagnosis of coronary artery disease[47]. Such unintentional depletions contribute to biases and complicate the interpretation of the proteomic results.

Figure 2 illustrates the analysis of an extracellular vesicle (EV) fraction enriched from plasma, digested using trypsin and measured by data independent acquisition with an Orbitrap Eclipse. This sample has a reduced dynamic range, compared to the whole plasma proteome making it an interesting avenue for biomarker discovery. The plasma vesicle fraction represents about 1-2% of the plasma proteome, is enriched in tissue derived proteins, and depleted in abundant plasma proteins. The total ion current (ions per second) from just the MS1 signal was $>10^{12}$, of which 46.4% of the current could be assigned to a peptide sequence using the fragment ion data. This current represented >5 billion ions of which 1.2 billion ions (24.1%) were assigned to peptide sequences – not counting the ions measured in the MS/MS spectra. To perform similarly, single-molecule methods like Illumina would analogously need to collect billions of reads from a mixture biochemically separated into 1000s of individual samples (~1 million reads per sample; see Figure 3b). The signal is normalized between flow cells to achieve counts that can be comparable between flow cells with ~24% of the reads being able to be mapped back to the reference genome. This plasma EV analysis was not sample limited and, thus, represents an analysis near the upper end of what can be achieved for the analysis of ions per analysis time.

Assuming that emerging polypeptide counting methods can achieve the current throughput of Illumina NovaSeq for DNA (4 billion reads for $10,000), their cost for analyzing a mammalian proteome would be much higher than the cost for MS analysis. This also suggests that single-molecule counting approaches must be at least 20x cheaper than Illumina sequencing to be cost effective when compared with $500 per LC-MS/MS analysis. Stated another way, LC-MS/MS is currently more efficient at counting peptides than next generation sequencing is at counting oligonucleotides.

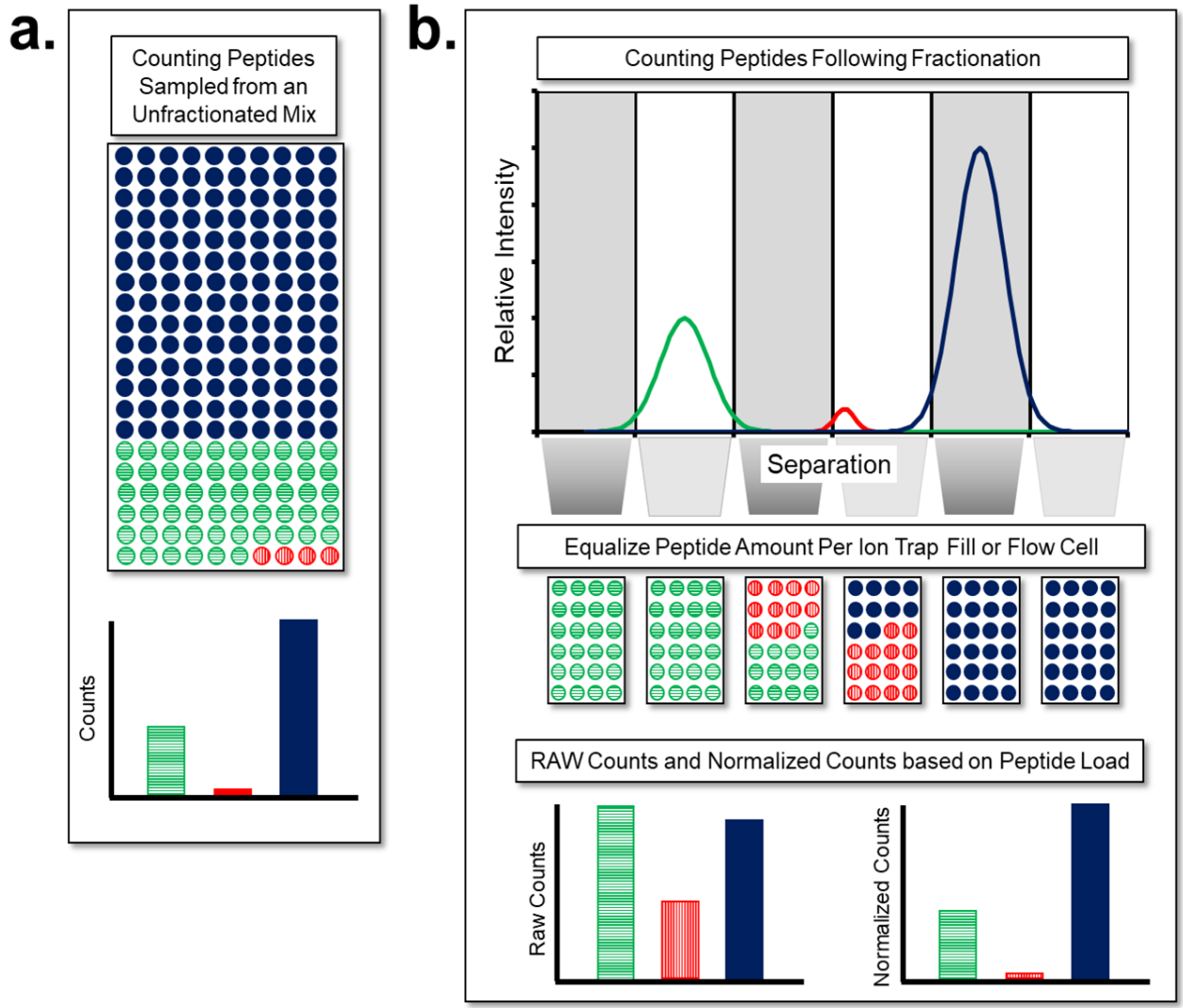

**Figure 3 | Fractionation prior to counting molecules improves the dynamic range in proteomics. a,** The dynamic range problem of the human proteome is far more extreme than that of transcriptomics. The enormous dynamic range of peptide abundances requires massive oversampling of the most abundant peptide (solid blue) in order to obtain counts for the least abundant peptide (vertical stripe red) **b,** Instead**,** LC-MS separates peptides biochemically, ionizes them, and samples the peptides at different times and with different spectra. While a mass spectrometer works in the gas phase, it is analogous to separating peptides/proteins prior to counting and applying normalization to make the quantities comparable between spectra or flow cells. This strategy significantly improves the counting statistics of low abundance molecules in the presence of high abundance molecules. In ion trap mass spectrometry, the normalization approach to optimize ions in each spectrum and adjust the signal by the variable fill time is known as automatic gain control (AGC).

**Scalability: the elephant in the room with single-molecule methods**

The sheer volume of protein molecules in a cell prompts a reality check - will single-molecule methods alone reach the required throughput to sufficiently sample the proteome? For single molecule counting methods to have the same coverage and breadth of the proteome as they do the transcriptome, they will need to have 10,000-30,000x more reads of similar quality as currently performed by RNA-seq. Thus, single molecule counting based methods will require technological advancements in nanopores, flow-cells and fluorescent detection methods that significantly exceed the capabilities of nucleotide single molecule counting methods.

A major factor that limits imaging based single-molecule sequencing is the density at which the molecules can be spaced, and the imaging strategies used to count the spatially resolved "reads" (we are assuming a 2D imaging plane in this discussion). The limit for the spatial density is constrained by the wavelength of light. Using fluorescence detection, the emission spectrum is in the 250-700 nm range (the actual theoretical resolution limit is about half the wavelength emitted). This provides a practical upper limit on planar molecular density of ~1 µm$^2$. Thus, assuming perfect measurement of reads and ideal spatial placement we can estimate the best-case scenario for the minimal flow cell area vs. number of reads: 1 million reads: 1 mm$^2$, 100 million reads: 1 cm$^2$, 10 billion reads: 10 cm$^2$, 1 trillion reads: 1 m$^2$. The area that needs to be imaged is limited by microscopy. These limits can be relaxed by super resolution imaging, but at the expense of decreased imaging speeds. Even with advances in widefield microscopy, there is a compromise between the field of view and the measured pixel size using a given charge coupled device (CCD) detector.

These estimates explain why 10 billion nucleotide reads is time consuming and expensive for sc-RNAseq analysis. Thus, the throughput of current nucleotide sequencing methods falls short of achieving the 400 billion reads needed for a full proteome analysis of a bulk sample at a similar coverage to that currently achieved on the transcriptome by RNA-seq.

Methods analyzing intact protein molecules, such as top-down MS[48] or single molecule methods that aim to count proteins[23], may be able to sample the proteome with less total counts. This is in contrast to peptide approaches that usually count multiple unique peptide sequences per proteoform. The difference between measuring intact proteoforms and peptides from the digestion of complex mixtures is analogous to the differences between short read RNA-seq and long read isoform sequencing. Intact protein analysis is further aided by recent methods for charge detection mass spectrometry (CDMS) where individual ion events can be measured[49,50].

A look at some alternative advanced single-molecule methods suggests a huge gap in throughput. The Pacific Biosciences Sequel II platform for genome sequencing can handle, at best, 10$^7$ molecules in each sequencing run, which takes a couple of days to complete[51]. The highest-throughput Oxford Nanopore Technology (ONT) platform, the Promethion, can run up to 48 flow cells at a time, providing an approximate maximum throughput of 5x10$^7$ molecules per run, which takes 1-2 days for data acquisition (signal processing time not included)[52]. These two examples are the most sophisticated single-molecule analyzers, and yet, the throughput offered is significantly short of the required throughput for analyzing protein mixtures on par with LC-MS. The high limit of 5x10$^7$ for single molecule technologies is no coincidence - these limitations are governed by physical limitations in scaling up device architecture for single-molecule interrogation, limitations in molecular turnover in the devices, as well as limitations in data acquisition and transfer rates. Taking the ONT pore sequencer and direct RNA sequencing as an example, 500 ng of input RNA contains about 10$^{12}$ mRNA molecules and only 10$^6$ of these are sampled in a MinION nanopore based flow cell. The vast discrepancy between input requirements and actual molecules analyzed (only 1 part per million is sampled!) is a testament to the intertwined limitations of single-molecule technologies: 500 ng ensures that molecules arrive to a nanoscale detector with minimal off-times, otherwise the sensor will be mostly vacated, and throughput will be compromised. In addition, the speed at which molecules pass through the pores cannot be too fast (typically 100 nm of polymer contour length per second), because the maximum measurement bandwidth of the electrical signal recording cannot exceed a few kHz due to data transfer speed and signal-to-noise limitations. These multiple constraints have set natural limits for single molecule processing, but there is no inherent reason for these to be hard limits. As flow cells are improved to enable analyses from smaller sample volumes, and/or strategies to deliver molecules more efficiently to the pores rather than rely on diffusion, one can imagine over 100-fold reductions in input requirements from >100 ng to <1 ng, at similar throughputs. Similarly, if one were to assume that data transfer and bandwidths would increase by ~100 fold over the next 5-7 years, one can expect transitioning from $10^3$ pores in a flowcell to $10^5$, which would boost the throughput 100-fold to about $5 \times 10^9$ molecules per run (1-2 days). It seems likely that these limitations will have to be overcome first for genomics/transcriptomics over the next 5-7 years, before single-molecule proteomics can be approached at scale using single-molecule tools.

**What limits LC-MS/MS and can the technology improve to sample the proteome?**

Most MS proteomics methods use a bottom-up strategy of digesting proteins to peptides to overcome the enormous physiochemical diversity of proteins in the cell. Overwhelmingly these methods make use of trypsin which produces peptides from proteins that have good cleavage specificity, are well suited for both reversed phase separations, produce mostly doubly and triply charged peptides, and fragment well because of the localization of a basic c-terminal residue and presence of a mobile proton. That said, not all tryptic peptides are well suited for LC-MS/MS, and because of this, proteins in complex mixtures are mainly identified through partial sequences. The sequence coverage of an identified protein varies between 10-100% (on average 30-50%) depending on the protein and the experiment. One approach to mitigate this limitation and maximize protein sequence coverage is to combine the results from different proteases with different specificity[53,54]. However, the increased sampling of ions derived from redundant peptides from the same proteins, while useful for improving coverage, comes at the expense of dynamic range as more ions must be sampled from additional peptides from abundant proteins before sampling ions from rare molecular species. To overcome the dynamic range problem alternative methods have been developed to minimize peptide coverage, capturing or depleting a subset of the peptides, while maximizing the different proteins sampled – this is analogous to exon capture[55], ChIP[56], or similar methods used in genomics prior to single molecule sequencing. Thus, there is a balance between maximizing coverage of individual proteins and the dynamic range of the proteins measured.

The major limiting factor in the sensitivity of LC-MS/MS methods is the electrospray process. Electrospray is the Nobel prize winning invention that is most commonly used for turning peptide molecules in solution into gas-phase ions[8]. However, if a molecule isn't converted to a gas-phase ion, it cannot be quantified with a mass spectrometer. Using electrospray, MS methods can quantify proteins present at 5,000 - 20,000 copies in the context of complex mammalian proteomes[12,57].

The number of ions sampled may be increased by using methods like multidimensional chromatography[14] or making multiple analyses using different portions of the mass range[58,59]. These approaches can significantly improve the depth of proteome coverage at the expense of increased analysis time. Thus, these gains in proteome coverage come at the cost of lower throughput and don't increase linearly with the time spent. For example, a 6x increase in time often only increases the number of peptides that can be measured by 2x – because the increased time is at least partially redundant with the peptides measured in prior fractions. Ultimately this comes at the expense of protein input material and significantly reduces the number of samples that can be measured. Thus a primary challenge is to achieve deep proteome coverage with smaller samples, such as single cells, and faster, thus enabling higher throughput[60,61].

Another way to improve LC-MS/MS is in the more efficient use of the ions that are generated. Currently, in most data independent acquisition methods, a single wide m/z range is isolated at once and the rest of the ion beam that isn't isolated is lost. Data dependent acquisition methods sample an even smaller fraction of the ion beam. With bulk samples, this means that only ~1/50th of the ion beam is currently being used as only one of 50 precursor windows is measured at once[62]. With single-cell samples, 3-4 windows are used and thus about ⅓ of all ions available to the MS instrument are analyzed[15] at the expense of limiting within spectrum selectivity. Methods like diaPASEF (parallel accumulation-serial fragmentation combined with data independent acquisition) offer potential to significantly increase the sampling of the peptide ion beam. Another important way to advance LC-MS/MS is to improve the computational methods that are used to assign peptide sequences to the ion current that is measured. Currently only ~15-50% of the measured ion current is assigned to peptide sequences[63]. Thus, an improvement in both the physical instrumentation for enhancing the sampling of the ion beam and computational methods for enhanced data interpretation could see a 50-75x improvement in the number of ions counted before LC-MS/MS becomes limited by the electrospray process. This improvement in ion counts will improve the relative measurement precision of the peptides measured, elevate low abundance species within the limit of detection, and enable measurements to be made in shorter time and with less material. We expect innovations in data acquisition and interpretation to enable quantification and sequence identification for a large fraction of the tens of thousands of peptide-like features detected in single cells, and thus substantially increase the depth of proteome coverage[63].

**What can emerging single molecule counting methods adopt from LC-MS/MS?**

Peptide quantification using LC-MS has evolved over the last several decades in ways that have improved our analyses of complex protein mixtures. Peptide ions are not counted one at a time but are aggregated, effectively compressing the signal from many peptide ions into a single measurement. This compression reduces time and minimizes the effect of abundant peptides on the counting precision of low abundant peptides – improving the dynamic range (Figure 3). However, the emphasis on generating and sorting 'like' ions constrains the choice of enzymes to produce peptides ideally suited for the respective method. Because tryptic peptides are ideally suited for LC-MS/MS doesn't mean it will be ideally suited for other methods. The conundrum is that reducing the bias by adding more distinct enzymes or nonspecific enzymes leads to more peptides with different sequences for each protein making it even harder to sample low abundance proteins in the presence of abundant proteins. Put simply, approaches to reduce these biases and increase sequence coverage in proteomics could push the field towards counting more ions from different peptide species – exacerbating the counting problem. Understanding the strengths and weaknesses of LC-MS as it has approached complex proteomes can perhaps constructively guide the emerging field of single-molecule proteomics. As advice to this budding field, consider the following.

*Fractionate:* Better to run many smaller counting experiments on fractionated samples than one very large counting experiment (Figure 2). If peptides or proteins are separated using an analytical method like liquid chromatography, electrophoresis, or affinity capture, the less abundant molecules will be enriched in certain fractions, resulting in a better representation of these peptides in the downstream detection/quantification processes. To make optimal use of this separation, methods equivalent to automatic gain control (AGC), as done with ion trap instruments[64], will need to be developed so that uniform fractions are fed into the flow cell for single-molecule readout. For example, each biochemical fraction can be diluted to the same concentration and equal quantities of the fractions loaded into many flow cells.

In addition to improving the dynamic range of the measurement, the use of a separation method based on a physicochemical property can be used to improve the sequence determination of the peptide or protein. In LC-MS/MS, the use of either predicted retention time or previously measured retention time is a powerful feature for the discrimination of correct and incorrect peptide detections[65–67]. This minimizes the FDR and improves sensitivity. Indeed, nanopore proteomics methods are making first steps in this direction[68].

The measurement of a signal across many points during a chromatographic separation also enables the integration of a chromatographic peak. Despite the unparalleled selectivity of LC-MS/MS measurements, there is often a background signal that complicates the quantitative linearity of the measurements. By integrating the peak along the separation, it is possible to perform a background subtraction, which improves quantitative accuracy.

***When there are many molecules to count, you will need to count many at a time.*** As mentioned above, to measure peptides using mass spectrometry from many billions of ions it became impractical to count ions one at a time in a realistic timescale. When done in a flow cell, single molecule counting methods will have to count so many molecules that they will likely either 1) exceed the density of the flow cell or 2) require a flow cell(s) with impractical physical dimensions. We hope to inspire new methods that are analogous to the switch in mass spectrometry from pulse counting (single molecule) to ion current measurement (each "read" will contain a variable quantity of many counts). Single molecule counting works great for transcripts because there are so few transcript molecules to count but the density can't scale easily 30,000x to extend to the dynamic range of the proteome.

***Overcoming biases.*** Arguably the most challenging aspect of proteomics is the massive physiochemical diversity of proteins in the cell. To overcome this vast diversity in solubility, embedded transmembrane domain containing proteins, size, combinatorial post-translational modification, ionization and fragmentation by mass spectrometry, presence of autoantibodies, or protein-protein interactions, most proteomics experiments take a bottom-up strategy for the analysis of complex mixtures by digesting proteins to peptides prior to analysis. Performing analyses on the peptide level greatly simplifies the physiochemical diversity of the analytes. In general, tryptic peptides are well matched for reversed phase chromatography, electrospray ionization, and tandem mass spectrometry. Methods for top-down proteomics have advanced enormously and have opened the door to characterizing proteoforms that are often ignored in understanding the function of the cell but these methods have greater constraints in their ability to analyze proteins with extremes in physiochemical properties[69].

Over the last two decades there have been massive improvements in nanoflow separations, electrospray ionization, transmission of ions from atmospheric pressure to vacuum, tandem mass spectrometry, and pipelined data acquisition that have resulted in sensitivities now approaching 10-50 zmol for peptides. However, one of the greatest challenges for single cell and low-input proteomics is the absorption of proteins and peptides to surfaces. In general, the sensitivity limits of proteomics samples have not been because of LC-MS/MS itself but the loss of sample to surfaces prior to entering the system. To solve these problems, there have been methods developed specifically to improve the recovery of protein from small numbers of cells using many strategies, including one-pot digestion[70,71], massively parallel sample preparation in surface droplets[72], addition of carrier proteins[73], and barcoding and combining samples using mass tags to spread losses between many samples[15].

Despite the potential sensitivity of emerging single molecule counting methods, these will need to overcome the same biochemical challenges of analyzing intact proteins, adsorptive losses to surfaces, variable enzyme digestion kinetics, and biases against certain peptide properties. While biases for sequencing peptides and proteins in flow-cells and nanopores will almost certainly be different than LC-MS/MS, the strategies for improving the recovery of peptides for entry into the instrument will largely be the same.

***Sample multiplexing:*** Peptides from multiple samples can be barcoded (e.g., by covalent chemical labels), subsequently mixed, and analyzed simultaneously. Sample multiplexing has helped increase the throughput of MS proteomics[11,74]. Analogous multiplexing methods are likely to be implemented by single-molecule methods to increase the number of samples analyzed as multiplexing is a powerful feature of single molecule DNA sequencing. Yet, multiplexing with single-molecule approaches spreads the counted molecules between many samples and thus reduces the number of molecules counted per sample, which results in shallower depth of proteome coverage and sequence completeness.

***Instrument companies historically focus on the bottom line before science.*** It is also important for new methods to have a clear fiscal return on investment. A couple of the new single molecule protein sequencing methods hope to convert peptide or protein sequences into DNA barcodes that can then be analyzed with traditional next generation sequencing technology[75]. However, as discussed above, the large number of protein molecules will require sequencing billions of molecules to obtain coverage of the proteome that can be obtained by LC-MS/MS[25]. Because this coverage can be obtained for ~$500 per analysis by LC-MS/MS and sequencing billions of DNA reads can cost ~$10,000, it would require Next Generation Sequencing companies to reduce their costs to ~5% their current rates. Without separation, a proteomic technology needs to count with high specificity about 1 billion intact protein molecules (or 20 billion peptides) for 100 USD to disrupt current LC-MS technologies. This price reduction would be a game changer for DNA sequencing and would further revolutionize genomics. However, it would require DNA sequencing companies to reduce their income from genomics applications to be financially competitive in the proteomics market. If they do this, then they will have done something that is rarely done in the proteomics field – minimize the financial return of existing products to be competitive in new high-risk areas.

**Summary**

Here we provided a perspective on the potential and challenges of scaling the use of single molecule counting methods to the analysis of the proteome. We use LC-MS based proteomics as a comparison by illustrating how many peptide molecules are counted in the gas-phase using standard mass spectrometry methods. This comparison will be useful for single molecule counting methods to use as a benchmark to obtain parity with LC-MS data. The challenges of analyzing the proteome by counting single peptide or protein molecules in a spatially resolved flow cell represents significant challenges over counting nucleotides – because of both the physiochemical complexity of proteins and the sheer greater number of proteins in the cell. To support innovation around these emerging methods we provide some suggestions learned by the LC-MS/MS based proteomics community.

**Acknowledgements**

This work was supported in part by National Institutes of Health grants U19 AG065156, R24 GM141156, F31 AG066318, an Allen Distinguished Investigator award through The Paul G. Allen Frontiers Group to N.S., a Seed Networks Award from CZI CZF2019-002424 to N.S., an R01 award from NIGMS R01GM144967, an R01 award from NHGRI R01HG10087 to M.W., the project 'International Centre for Cancer Vaccine Science' that is carried out within the International Agendas Programme of the Foundation for Polish Science co-financed by the European Union under the European Regional Development Fund. We thank the PL-Grid and CI-TASK Infrastructure, Poland, for providing their hardware and software resources. This work is supported by "Knowledge At the Tip of Your fingers: Clinical Knowledge for Humanity" (KATY) project funded from the European Union's Horizon 2020 research and innovation program under grant agreement No. 101017453. The authors would like to acknowledge the helpful discussions with Edward Marcotte and members of the Alfaro, MacCoss, and Slavov labs. MJM appreciates the constructive feedback provided by UW Genome Sciences faculty.